\documentclass[aps,prb,reprint,amssymb,amsmath,superscriptaddress]{revtex4-1}
\usepackage{graphicx}
\usepackage{epstopdf}

\usepackage[english]{babel}
\bibpunct{[}{]}{,}{n}{}{}
\begin{document}

\title{Anisotropy of in-plane  $\textsl{g}$-factor of electrons in HgTe quantum wells}

\author{G. M.~Minkov}

\affiliation{School of Natural Sciences and Mathematics, Ural Federal University,
620002 Ekaterinburg, Russia}

\affiliation{M.~N.~Miheev Institute of Metal Physics of Ural Branch of
Russian Academy of Sciences, 620137 Ekaterinburg, Russia}

\author{V.\,Ya.~Aleshkin}
\affiliation{Institute for Physics of Microstructures  RAS, 603087 Nizhny Novgorod, Russia}

\affiliation{Lobachevsky University of Nizhny Novgorod, 603950
Nizhny Novgorod, Russia}

\author{O.\,E.~Rut}
\affiliation{School of Natural Sciences and Mathematics, Ural Federal University,
620002 Ekaterinburg, Russia}

\author{A.\,A.~Sherstobitov}

\affiliation{School of Natural Sciences and Mathematics, Ural Federal University,
620002 Ekaterinburg, Russia}

\affiliation{M.~N.~Miheev Institute of Metal Physics of Ural Branch of
Russian Academy of Sciences, 620137 Ekaterinburg, Russia}

\author{S.\,A.~Dvoretski}

\affiliation{Institute of Semiconductor Physics RAS, 630090
Novosibirsk, Russia}

\author{N.\,N.~Mikhailov}

\affiliation{Institute of Semiconductor Physics RAS, 630090
Novosibirsk, Russia}
\affiliation{Novosibirsk State University, Novosibirsk 630090, Russia}

\author{A.~V.~Germanenko}

\affiliation{School of Natural Sciences and Mathematics, Ural Federal University,
620002 Ekaterinburg, Russia}

\date{\today}

\begin{abstract}
The results of experimental studies of the Shubnikov–de Haas (SdH) effect in the (013)-HgTe/Hg$_{1-x}$Cd$_x$Te quantum wells (QWs) of electron type of conductivity both with normal and inverted energy spectrum are reported. Comprehensive analysis of the SdH oscillations measured for the different orientations of magnetic field relative to the quantum well plane and crystallographic exes allows us to investigate the anisotropy of the Zeeman effect. For the QWs with inverted spectrum, it has been shown that the ratio of the spin splitting to the orbital one is strongly dependent not only on  the orientation of the magnetic field relative to the QW plane but also on the orientation of the in-plane magnetic field component relative to crystallographic axes laying in the QW plane that implies the strong anisotropy of in-plane $\textsl{g}$-factor. In the QW with normal spectrum,  this ratio strongly depends on the angle between the magnetic field and the normal to the QW plane and reveals a very slight anisotropy in the QW plane. To interpret the data, the Landau levels in the tilted magnetic field are calculated within the framework of  four-band \emph{kP} model. It is shown that the experimental results can be quantitatively described only with taking into account the interface inversion asymmetry.

\end{abstract}


\maketitle

\section{Introduction}
\label{sec:intr}

Spin-dependent effects in transport, tunneling, optical phenomena are interesting and important not only for understanding the role of these effects in all phenomena \cite{Dyak17}, but also for possible application. These effects are largely determined by the $\textsl{g}$-factor and its anisotropy, that is by its dependence on the direction of the magnetic field relative to the two-dimensional plane and crystallographic axes. The $\textsl{g}$-factor anisotropy can be very strong and important in 2D structures based on materials with a large spin-orbit interaction, a complex spectrum, and in the structures grown on substrates with a low-symmetric surface. The HgTe quantum wells belong to such a type of structures.

The energy spectrum of HgTe quantum wells (QWs) is complicated and depends strongly on the quantum well width ($d$). For $d <d_c\simeq 6.3$~nm, the conduction band is formed from electron states and the states of the light hole \cite{Gerchikov90,Zhang01,Novik05,Bernevig06,ZholudevPhD}. This type of the spectrum is named normal. At $d > d_c$, the conduction band is formed from heavy-hole states and such type of the spectrum is named ``inverted''. At $d=d_c$, the linear in quasimomentum ($k$) gapless spectrum  is realised.
Experimentally, the  $\textsl{g}$-factor  and its anisotropy was investigated  in the structures both with normal and with ``inverted'' spectrum grown on substrates of different orientations \cite{Zhang04, Molenkamp2014,Yak12,Kozlov14,Bovkun15}.  In all the cases it was assumed that the in-plane $\textsl{g}$-factor ($\textsl{g}_{\parallel}$) is isotropic.

In this paper, we study the angle dependences of the  amplitude of the Shubnikov-de Haas (SdH) oscillations in tilted magnetic fields in (013)-HgTe QWs with both types of energy spectrum.  We show that the ratio of the spin to orbit splitting  is strongly anisotropic and this anisotropy is strongly different for QWs with $d>d_c$ and $d<d_c$. Especially, it concerns the anisotropy of  in-plane $\textsl{g}$-factor.  The paper is organized as follows. The samples and experimental conditions are described in the next section. The experimental results and their analysis for the QW of $d=10$~nm with ``inverted'' energy spectrum are presented in Sec.~\ref{sec:exp}. The surprising finding is that the oscillation picture in the tilted magnetic field is strongly different for the two cases when the in-plane component changes its direction on the angle of $180^\circ$. It points to the strong anisotropy of the in-plane $\textsl{g}$-factor. In Sec.~\ref{sec:th} we describe theoretical model allowing us to calculate the spectrum of the Landau levels (LLS)  in the tilted magnetic field.  Comparison of the data for the QW with $d>d_c$ with theoretical results is performed  in Sections~\ref{sec:th0} and \ref{secV}. The data obtained for QW with $d<d_c$ are inspected and analysed in Sec.~\ref{secVI}. Section~\ref{sec:conc} is devoted to the conclusions.

\section{Experimental}
\label{sec:expdet}

Our samples with the HgTe quantum wells  were realized on the basis of
HgTe/Hg$_{1-x}$Cd$_{x}$Te heterostructures grown by the
molecular beam epitaxy on a GaAs substrate with the (013) surface
orientations \cite{Mikhailov06}.  The samples were mesa etched into standard Hall bars of
$0.5$~mm  width and the distance between the potential probes was
$0.5$~mm. To change and control the carrier density in the quantum
well, the field-effect transistors were fabricated with parylene as an
insulator and aluminium as a gate electrode. For each heterostructure,
several samples were fabricated and studied. The parameters of the structures are presented in the Table~\ref{tab1}.

\begin{table}
\caption{The parameters of  heterostructures under study}
\label{tab1}
\begin{ruledtabular}
\begin{tabular}{ccccccc}
number & structure & $d$ (nm)&  $x$& $n(V_g=0)$ (cm$^{-2}$) \\
\colrule
  1& 150224 & 10.0  &0.52    &  $1.15\times 10^{11}$      \\
  2& 150220 & 4.6  &0.54   &  $1.60\times 10^{11}$ \\
\end{tabular}
\end{ruledtabular}
\end{table}

All measurements were carried out using the DC technique  at $T=4.2$~K within magnetic field range $(0-6)$~T.

\begin{figure}
\includegraphics[width=0.9\linewidth,clip=true]{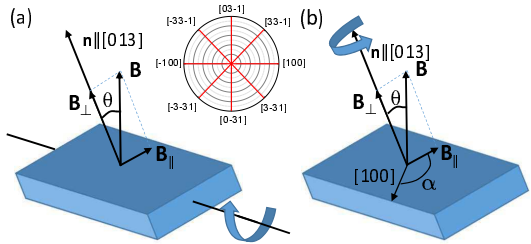}
\caption{(Color online) Two configurations in which the angle dependences  were measured. The inset shows the crystallographic exes laying in the (013) QW plane. }
\label{F0}
\end{figure}

The ratio of the spin splitting ($\Delta_s$) to the orbital one ($\Delta_o$) $X=\Delta_s/\Delta_o$ was obtained by means of modified coincidence method \cite{Fang1968,Stud05,Minkov17-1,Kurganova11}.  The measurements were taken in two configurations which are shown in Fig.~\ref{F0}. In the first configuration, which is widely used be experimentalists, the rotation axis is perpendicular to the magnetic field and lies in the 2D plane [Fig.~\ref{F0}(a)].  In the second configuration, the rotator and the sample  are oriented in such a way that the axis of rotation is normal to the 2D plane and tilted relative to the magnetic field [see Fig.~\ref{F0}(b)].  This allowed us to investigate the in-plane anisotropy of spin-to-orbit splitting ratio.

\section{Expreimental results}
\label{sec:exp}

Let us begin with analysis of the results obtained in the first configuration. The rotation on the angle $\theta$, as shown in Fig.~\ref{F0}(a), changes the ratio between the normal and in-plane components ($B_\perp$ and $B_\parallel$, respectively)  of magnetic field; $B_\perp=B\, \cos{\theta}$, $B_\parallel=B\,\sin{\theta}$, where $B$ is the total magnetic field. If we assume that the spin splitting is proportional to the total magnetic field $\Delta_s\propto B$ and the orbital splitting is proportional to the normal component $\Delta_o\propto B_{perp}$, the angle dependence of the oscillation amplitude in the low magnetic fields in which the spin splitting of oscillations is not resolved looks as follows \cite{Stud05,Germ10,Minkov17-1}
\begin{equation}
\frac{A(\theta)}{A(0)} = \frac{\cos[\pi X(\theta)]}{\cos[\pi X(0)]},
\label{eq1}
\end{equation}
where
\begin{equation}
X(\theta)=\frac{\Delta_s}{\Delta_o(\theta)}=\frac{\textsl{g}\mu_B B}{(e\hbar B_\perp/m)}
\label{eq10}
\end{equation}
with $\mu _B$ as the Bohr magneton and $m$ as effective mass.
If the $\textsl{g}$-factor is  anisotropic, the spin splitting of the Landau levels (LL) becomes angle dependent as well. In the simplest case it can be written as follows
\begin{equation}
\textsl{g}(\theta)=\sqrt{\textsl{g}_\perp^2 \cos(\theta)^2+\textsl{g}_\parallel^2 \sin(\theta)^2}.
\label{eq2}
\end{equation}
and then
\begin{equation}
X(\theta)=\frac{\Delta_s(\theta)}{\Delta_o(\theta)}=\frac{\sqrt{\textsl{g}_\perp^2 \cos(\theta)^2+\textsl{g}_\parallel^2 \sin(\theta)^2}\mu_B B}{(e\hbar B_\perp/m)}.
\label{eq2s}
\end{equation}
So, the measurements of the SdH oscillation amplitude at fixed $B_\perp$ as a function of tilt angle $\theta$ give, in principle, a possibility to find the ratio of the spin splitting to the orbital one and to obtain the $\textsl{g}$-factor value.

This method is valid when: (i) $B_\perp$ is significantly less than the field of the onset of the quantum Hall effect (QHE); (ii) the amplitude of the oscillations is small so that the oscillations of the Fermi energy are negligible; (iii) the SdH oscillations are spin-unsplit.

\begin{figure}
\includegraphics[width=1\linewidth,clip=true]{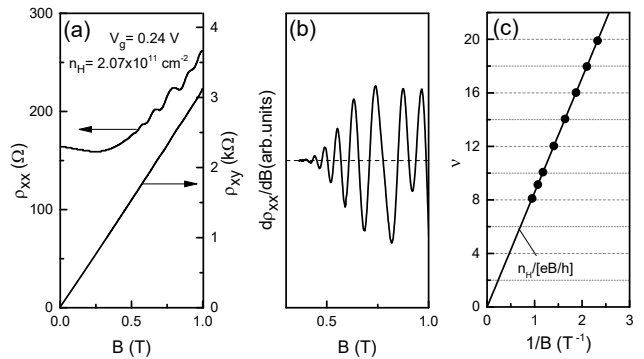}
\caption{(Color online) The magnetic field dependences of $\rho_{xx}$, $\rho_{xy}$   (a) and the oscillations of $d\rho_{xx}/dB$ (b). (c) -- The dependence of the filling factor $\nu$ on positions of the $\rho_{xx}$ minima in the reciprocal magnetic field. $\theta=0$. }
\label{F1}
\end{figure}

In this section we analyze the results obtained for the structures 1 with ``inverted'' spectrum ($d=10$~nm).  In Fig.~\ref{F1}(a), we  present  the magnetic field dependences of $\rho_{xx}$ and $\rho_{xy}$  measured at $\theta=0$  for the electron density $n = 2.07\times 10^{11}$~cm$^{-2}$. As seen  the amplitude of $\rho_{xx}$ oscillations  is less than $10$ percent and $\rho_{xy}$ linearly depend on $B$  (the steps of OHE are absent). This means that the oscillations of the Fermi energy within this magnetic field range can be neglected. The electron density found from the period of oscillations in $B<0.7$~T under assumption that the Landau levels are two-fold degenerate, coincides with the Hall density $n_H=-1/eR_H$.
So, the conditions of applicability of Eq.~(\ref{eq1}) are met.

\begin{figure}
\includegraphics[width=\linewidth,clip=true]{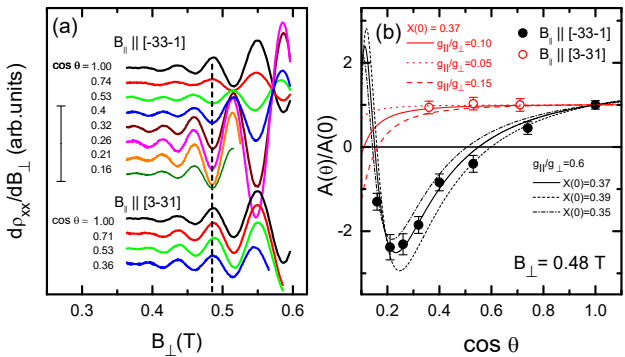}
\caption{(Color online) (a)  -- The dependences $d\rho_{xx}(B_\perp)/dB_\perp$ for some angles $\theta$, when the in-plane component  appears in the directions $[\bar{3}3\bar{1}]$  and $[3\bar{3}1]$, (b) -- The angle dependences of the oscillation amplitude measured at $B_\perp=0.48$~T (symbols) and calculated as described in the text (curves). The scale shown in the panel (a) by the vertical bars is the same as in Fig.~\ref{F5}(a). Structure 1, $n=2.07\times10^{11}$~cm$^{-2}$. }
\label{F2}
\end{figure}

Let us now inspect the SdH oscillations in more detail. To remove the monotonic part we plot in Fig.~\ref{F0}(b) the
dependence $d\rho_{xx}(B)/dB$. One can see that the oscillations which appear at $B\simeq 0.3$~T start to split at $B\simeq0.7$~T. To understand the degeneracy of the Landau levels with which the observed oscillations are associated, it is instructive to consider the filling factor $\nu=n_H/(eB_{min}/h)$ plotted against $1/ B_{min}$, where $B_{min}$ is magnetic fields at which the minima in $\rho_{xx}(B)$ are observed. Such a dependence is represented in  Fig.~\ref{F1}(c).  It is evident that  for $1/B>1.2$~T$^{-1}$, $\nu>10$ the filing factor changes by two and takes the even values therewith.
This is a clear indication of  the fact that the oscillations for these $\nu$ are associated with the  pairwise merged Landau levels which are two spin sublevels with the same orbital number. Thus we infer that the spin splitting $\Delta_s$ is less than half the orbital splitting $\Delta_o$, i.e.,  $X(0) =\Delta_s/\Delta_o <0.5$.

Let us now consider the oscillations in tilted magnetic fields. The dependences of the $d\rho_{xx}/dB_\perp$ on the normal to 2D plane magnetic field $B_\perp$ for the case when the in-plane component appears in directions
$B_\parallel \upuparrows[\bar{3}3\bar{1}]$ and
$B_\parallel \upuparrows[3\bar{3}1]$  are shown in Fig.~\ref{F2}(a) for some tilt angles. It is seen that the positions of the oscillations in $B_\perp$  are independent of $\theta$ within the experimental accuracy, while  the amplitude of the oscillations varies significantly.

The angle dependences of the normalized oscillation amplitude $A(\theta)/ A(0)$ at $B_\perp=0.48$~T  are represented in Fig.~\ref{F2}(b) \footnote{To determine the oscillation amplitude, the experimental curves were fitted by the Lifshits-Kosevich  formula \cite{LifKos55} as described in Ref.~\cite{Minkov17}}.
The negative sign of  $A(\theta)/ A(0)$ corresponds to jump of the oscillation phase on $\pi$. Particularly striking is that the angle dependences of oscillation amplitudes are drastically different for $B_\parallel \upuparrows[\bar{3}3\bar{1}]$ and
$B_\parallel \upuparrows[3\bar{3}1]$. The amplitude immediately decreases with the $\cos{\theta}$ decrease when $B_\parallel \upuparrows[\bar{3}3\bar{1}]$
and does not practically depend on $\theta$ when $B_\parallel \upuparrows[3\bar{3}1]$ within the actual $\theta$ range.
All this  indicates that the in-plane $\textsl{g}$-factor $\textsl{g}_\parallel$ differs dramatically  for two opposite crystallographic directions $[\bar{3}3\bar{1}]$ and $[3\bar{3}1]$.

To determine the $X(0)$ values  and $\textsl{g}$-factor anisotropy $\textsl{g}_{[\bar{3}3\bar{1}]}/\textsl{g}_\perp$ and $\textsl{g}_{[3\bar{3}1]}/\textsl{g}_\perp$, the $A(\theta)/A(0)$~vs~$\theta$ data  in Fig.~\ref{F2}(b) were fitted  by Eqs.~(\ref{eq1}) and (\ref{eq2s}) with the use of $X(0)$ and $\textsl{g}_{[\bar{3}3\bar{1}]}/\textsl{g}_\perp$ (for the solid circles) and $\textsl{g}_{[3\bar{3}1]}/\textsl{g}_\perp$ (for the open ones) as the fitting parameters. The results of the best fit are shown  in Fig.~\ref{F2}(b) by the solid curves
\footnote{It should be noted that the $X(0)$ and $\textsl{g}_\parallel/\textsl{g}_\perp$ values can be directly  obtained from the data without the fitting procedure if  the experimental   $A(\theta)/A(0)$~vs~$\theta$ plot has one or more extrema at some angle(-s) $\theta=\theta^\star$. As follows from Eqs.~(\ref{eq1}) and (\ref{eq2s})  the experimental values of $X(0)$ and $\textsl{g}_\parallel/\textsl{g}_\perp$ in such a case can be obtained as
$$
X^\text{exp}(0)= \frac{1}{\pi} \arccos{\left[-\frac{A(0)}{A(\theta^\star)}\right] }
$$
and
$$
\left(\frac{\textsl{g}_\parallel}{\textsl{g}_\perp}\right)^\text{exp}=\frac{1}{(\tan{\theta^\star})^2}\sqrt{\frac{1}{[X^\text{exp}(0)]^{2}}-1},$$
respectively}. It is seen that the data for both directions $[\bar{3}3\bar{1}]$ and $[3\bar{3}1]$ are well fitted by Eqs.~(\ref{eq1}) and (\ref{eq2s}), that allows us to obtain the $\Delta_s/\Delta_o$ value for $\theta=0$, $X(0)=0.37\pm 0.02$, and the values of  $\textsl{g}_{[\bar{3}3\bar{1}]}/\textsl{g}_\perp$ and $\textsl{g}_{[3\bar{3}1]}/\textsl{g}_\perp$
equal to $0.6\pm 0.1$ and  $0.10\pm0.05$, respectively.

Thus, the above analysis shows that  the ratio of the spin splitting to the orbital one for $\theta = 0$ is $X(0) = 0.37\pm 0.02$. Therewith $\textsl{g}_\parallel$ depends strongly on the crystallographical directions.

Let us now compare the experimentally found value of $X(0)$ with theoretical one. To do it  we calculate the energies of LLs  in a magnetic field of arbitrary orientation.

\section{ The Landau levels in tilted magnetic field}
\label{sec:th}

Let us choose the vector potential so that only the components lying in the
plane of the quantum well are nonzero:
\begin{equation}
A_x=A_x'-B_yz,\quad A_y=A_y'-B_xz
\end{equation}
The vector potential components  with strokes describe
the magnetic field along the $z$ axis:
\begin{equation}
H_z=\frac{\partial A_y'}{\partial x}-\frac{\partial A_x'}{\partial y}
\end{equation}
Let introduce creation and annihilation operators:
\begin{eqnarray}
a^+&=&\frac{\lambda}{\sqrt{2}}\left(k_x+ik_y\right)=\frac{\lambda}{\sqrt{2}}k_+,\quad \nonumber \\
a&=&\frac{\lambda}{\sqrt{2}}\left(k_x-ik_y\right)=\frac{\lambda}{\sqrt{2}}k_-
\end{eqnarray}
where
\begin{eqnarray}
k_x&=&-i\frac{\partial}{\partial x}+\frac{eA_x'}{\hbar c}, \nonumber  \\
k_y&=&-i\frac{\partial}{\partial y}+\frac{eA_y'}{\hbar c}, \\
\lambda&=&\sqrt{\frac{\hbar }{|eB_z|}}. \nonumber
\end{eqnarray}
The
operators $k_x$ and $k_y$ satisfy the following commutation relation:
\begin{equation}
[k_x,k_y]=-i\frac{eB_z}{\hbar }
\end{equation}
and therefore
\begin{equation}
[a,a^+]=\frac{eB_z}{|eB_z|}
\end{equation}
Further we supply that $B_z>0$, so $[a,a^+]=1$. To calculate the energy spectrum we have used the $8\times 8$ Kane Hamiltonian which takes exactly into account interactions between the bands $\Gamma_6$, $\Gamma_8$, and $\Gamma_7$.  The interactions with the other remote bands are taken into account as the second-order perturbations. An explicit form of the Hamiltonian  is given in Refs.~\cite{ZholudevPhD,Zholudev12} for the quantum well grown on the (013) plane. To incorporate the magnetic field, the following substitutions have been made
\begin{eqnarray}
k_+&\rightarrow& a^+i\frac{\sqrt{2}}{\lambda}\frac{ezB_+}{\hbar },\nonumber \\
k_-&\rightarrow& a\,i\frac{\sqrt{2}}{\lambda}\frac{ezB_-}{\hbar },
\end{eqnarray}
where $B_\pm =B_x\pm iB_y$. Moreover, to take into account the Zeeman effect, the following term has been added to the Hamiltonian
\begin{equation}
H_Z=\frac{e\hbar}{m_0}\left(
\begin{array}{ccc}
H_{cc}&0\\
0&H_{vv}
\end{array}
\right),
\end{equation}
where
\begin{equation}
H_{cc}=\frac{1}{2}\left(
\begin{array}{cc}
B_z&B_-\\
B_+&-B_z\\
\end{array}
\right),
\
\end{equation}
and
\begin{widetext}

\begin{equation}
H_{vv}=\left(
\begin{array}{cccccc}
-\frac{3\kappa}{2}B_z&-\frac{\sqrt{3}\kappa}{2}B_-&0&0&\frac{\sqrt{6}(\kappa+1)}{4}B_-&0\\
-\frac{\sqrt{3}\kappa}{2}B_+&-\frac{\kappa}{2}B_z&-\kappa B_-&0&-\frac{(\kappa +1)}{\sqrt{2}}B_z&\frac{(\kappa +1)}{2\sqrt{2}}B_-\\
0&-\kappa B_+&\frac{\kappa }{2}B_z&-\frac{\sqrt{3}\kappa }{2}B_-&-\frac{(\kappa +1)}{2\sqrt{2}}B_+&-\frac{(\kappa +1)}{\sqrt{2}}B_z\\
0&0&-\frac{\sqrt{3}\kappa }{2}B_+&\frac{3\kappa }{2}B_z&0&-\frac{\sqrt{6}(\kappa +1)}{4}B_+\\
\frac{\sqrt{6}(\kappa +1)}{4}B_+&-\frac{(\kappa +1)}{\sqrt{2}}B_z&-\frac{(\kappa +1)}{2\sqrt{2}}B_-&0&-\left(\kappa +\frac{1}{2}\right)B_z&-\left(\kappa +\frac{1}{2}\right)B_-\\
0&\frac{(\kappa +1)}{2\sqrt{2}}B_+&-\frac{(\kappa +1)}{\sqrt{2}}B_z&-\frac{\sqrt{6}(\kappa +1)}{4}B_-&-\left(\kappa +\frac{1}{2}\right)B_+&-\left(\kappa +\frac{1}{2}\right)B_z\\
\end{array}
\right).
\end{equation}
\end{widetext}

In order to find the eigenvalues and eigenfunctions of the total
Hamiltonian, we divided it into  axially symmetric and axially
asymmetric parts. On the first step we find eigenvalues and
eigenfunctions of the axially symmetric part. To do it we use procedure
described in Refs.~\cite{ZholudevPhD,Minkov17}.
On the second step we use these eigenfunctions as a basis for expansion of
wave function of the total Hamiltonian. Then the total Hamiltonian was
represented as a matrix on the basis of these eigenfunctions, which
eigenvalues and eigenfunctions represent the solution of our problem. Note that eigenfunctions
of the axially symmetric Hamiltonian have two quantum numbers: number of
the Landau level and subband number.  In expansion we usually used  $30-50$ Landau levels and $10-15$ subbands for the calculation of electron states in our structures within the interval of magnetic field used in our experiments. Further increase of the Landau and subband numbers does not practically change the calculation results.

The parameters used in the calculations are listed in the Table~\ref{tab2}. Values for deformation potentials $a_c$, $a_v$, $b$, $d$ and elastic constants $C_{ij}$ were taken from Ref.~\cite{Takita79}; parameter $B_8$ is from Ref.~\cite{Wink02-1}; other parameters are from Ref.~\cite{Novik05}. All the values were assumed linearly dependent on  $x$ excepting $E_g(x)$ which is calculated in accordance with Refs.~\cite{Becker00,Laurenti90}.
The valence band offset value equal to the difference of HgTe and CdTe valence band maximum energy $E_v(\text{HgTe})-E_v(\text{CdTe})$ is $570$~meV \cite{Becker00}.

\begin{table}
\caption{The parameters used in the calculations}
\vspace{5mm}
\label{tab2}
\begin{tabular}{lcc}
\hline
parameter & CdTe & HgTe  \\
\hline
$E_g$ (eV) & 1.606 & -0.303 \\
$E_v$ (eV) & -0.57 & 0\\
$\Delta$ (eV) & 0.91 & 1.08 \\
$F$ & -0.09 & 0 \\
$E_p$ (eV) & 18.8 & 18.8 \\
$\gamma_1$ & 1.47 & 4.1 \\
$\gamma_2$ & -0.28 & 0.5\\
$\gamma_3$ & 0.03 &1.3 \\
$\kappa$ & -1.31 & -0.4 \\
$B_8$ (eV$\cdot$\AA$^2$) & -22.41 & – \\
$a$ (\AA) &  6.48 & 6.46 \\
$a_c$ (eV) & -2.925 & -2.380 \\
$a_v$ (eV) & 0 & 1.31 \\
$b$ (eV) & -1.2 & -1.5 \\
$d$ (eV) & -5.4 & -2.5 \\
$C_{11}$ ($10^{11}$~din/cm$^2$) & 5.62 & 5.92\\
$C_{12}$ ($10^{11}$~din/cm$^2$) & 3.94 & 4.14 \\
$C_{44}$ ($10^{11}$~din/cm$^2$) & 2.06 & 2.19\\
\hline
\end{tabular}
\end{table}

\section{ Comparison between theory and experiment for $\theta=0$	
}

\label{sec:th0}

Firstly we have calculated  the Fermi energy ($E_F$) which is equal to $27$~meV for $n=2.07\times10^{11}$~cm$^{-2}$ that corresponds to the data represented in Figs.~\ref{F1} and \ref{F2}. Then we have found the cyclotron energy and the spin splitting at the energy close to the Fermi energy as the differences  $\Delta_o=E_{N+2}-E_{N}$ and $\Delta_s=E_{2N}-E_{2N-1}$, respectively, where $N$ is the number of the LLs numbered in a row starting with $N=1$. For the actual case of $n=2.07\times10^{11}$~cm$^{-2}$ and $B_\perp=0.5$~T  (see Fig.~\ref{F2}), the Landau levels laying close to the Fermi level have the numbers $N=16-18$ and this estimate gives $\Delta_o\simeq 2.68$~meV, $\Delta_s\simeq1.55$~meV that corresponds to the following values of effective mass and spin-to-orbit splitting ratio: $m=0.0216\, m_0$ and $X^\text{calc}(0)=0.58$. Recall that $X^\text{exp}(0)\simeq 0.37$ (see Section~\ref{sec:exp}). The difference between $X^\text{calc}(0)$ and  $X^\text{exp}(0)$  is radical not only quantitatively but qualitatively.
Really, when $X(0)> 0.5$, the different spin sublevels of the neighboring LLs with different orbital numbers should merged in low magnetic fields and  the only odd minima in $\rho_{xx}(B)$ should be observed in this case. When $X(0)<0.5$ the different spin  sublevels  with one and the same orbital number   are merged and the only even minima in $\rho_{xx}(B)$ should be observed. As Fig.~\ref{F1}(c) shows, at $B_\perp <0.7$~T we observe even minima that accords well with $X^\text{exp}(0)=0.37<0.5$, while the theory predicts $X^\text{calc}(0)=0.58>0.5$ which should lead to observation of the odd minima.

We assume that two factors can be responsible for such a difference between $X^\text{exp}(0)$ and $X^\text{calc}(0)$. Let us consider the first one.

\begin{figure}
\includegraphics[width=\linewidth,clip=true]{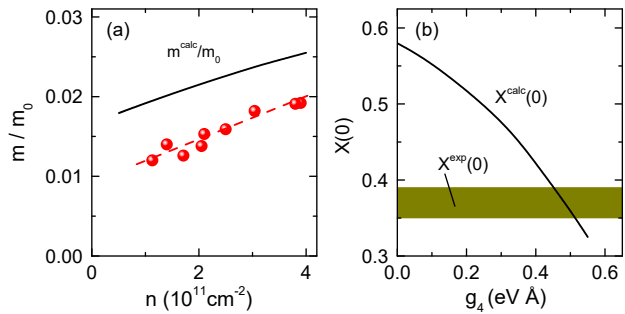}
\caption{(Color online) (a) -- The calculated (solid curve) and experimental (symbols) electron density dependences of the effective mass.
The dashed line is a guide to the eye.   (b) -- The calculated ratio of the spin splitting to orbital one $X(0)$ as a function of parameter $g_4$ which controls the contribution of interface inversion asymmetry. }
\label{F4}
\end{figure}

Experimentally, the used method gives the ratio of the spin splitting to the orbital one. The orbital splitting $\Delta_o=\hbar \omega_c=\hbar eB/m$  is determined by the effective mass at the Fermi energy. The  studies of $m$ as a function of QW width and electron density \cite{Minkov18} show that the $m$ value in the  structure under consideration is equal to $(0.015\pm0.002)\,m_0$ at $n=2.07\times10^{11}$~cm$^{-2}$, while the above calculation  gives $m^\text{calc}=0.0216\, m_0$ [see Fig.~\ref{F4}(a)]. There was supposed in Ref.~\cite{Minkov18} that such a difference between theory and experiment may result from many-body effects which are not taken into account in the theory used. If one supposes that the many-body effects lead only to decrease in the effective mass  but do not change the $\textsl{g}$-factor, one should correct the calculated value of $X^\text{calc}(0)$ by the following way  $X^\text{corr}(0)=X^\text{calc}(0)\times m^\text{exp}/m^\text{calc}=0.58\times m^\text{exp}/m^\text{calc}\simeq0.4\pm 0.03$. As seen  such a correction gives well coincidence with the experimental value $X^\text{exp}(0)=0.37\pm 0.02$.

Another reason for the difference between $X^\text{exp}(0)$ and $X^\text{calc}(0)$ may be the result of the fact that the interface inversion asymmetry (IIA) was not taken into account yet in the calculation described in Section~\ref{sec:th}. To take into account IIA we used an additional term in the Hamiltonian, which is suggested by Ivchenko \cite{Ivch05} (for more details, see also  Ref.~\cite{Minkov17}). The results of the calculations  are shown in Fig.~\ref{F4}(b) where $X^\text{calc}(0)$ is plotted as a function of the value of the parameter $g_4$, which controls the IIA contribution.  It is seen that taking into account only IIA with the values of the parameter $g_4 = (0.45-0.53)$~eV\,\AA\,  also gives a good agreement with the experimental value $X^\text{exp}(0)$.

Thus, comparison of only $X^\text{exp}(0)$ with the theoretical results  does not give an unambiguous answer to the question which factor, mass renormalization or interface inversion asymmetry, gives the main contribution to the difference between theory and experiment.

\section{  In-plane anisotropy of $\textsl{g}$-factor, $d>d_c$ }
\label{secV}

For a detailed experimental study of the in-plane anisotropy of the Zeeman splitting we measured the SdH oscillations in  the configuration shown in Fig.~\ref{F0}(b). The rotator and the sample were set in such a way that the axis of rotation was normal to two-dimensional gas and tilted relative to the magnetic field by the angle $\theta$.  This angle determines the ratio of the normal to in-plane component of the magnetic field.  At fixed magnetic field $B$, the rotation in this case changes the direction of $B_\parallel$ with respect to crystallographic axis but does not change the ratio  $B_\parallel/B_\perp=\tan{\theta}$.

\begin{figure}
\includegraphics[width=\linewidth,clip=true]{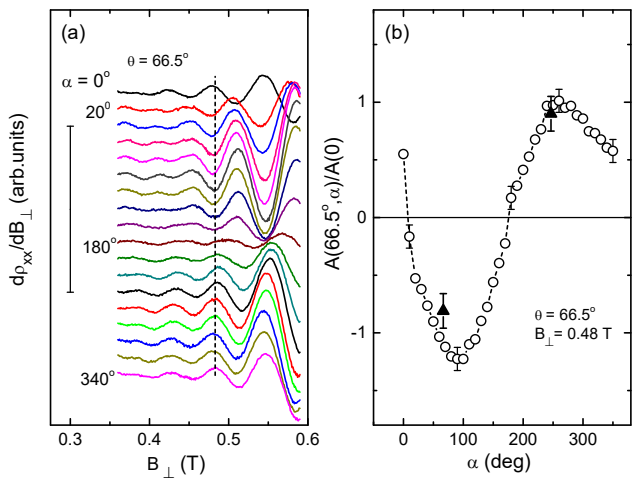}
\caption{(Color online) (a) -- The $B_\perp$ dependences of  $d\rho_{xx}/dB_\perp$  for the different   angles $\alpha$  measured  at $\theta=66.5^\circ$. (b) -- The normalized oscillation amplitudes $A(66.5^\circ, \alpha)/A(0)$ plotted against the  angle $\alpha$. The triangle shows  $A(66.5^\circ, 0)/A(0)$  measured in the  first configuration [see Fig.~\ref{F2}(b)]. The bar in panel (a) is the same as shown in Fig.~\ref{F2}(a).
}
\label{F5}
\end{figure}

\begin{figure}
\includegraphics[width=\linewidth,clip=true]{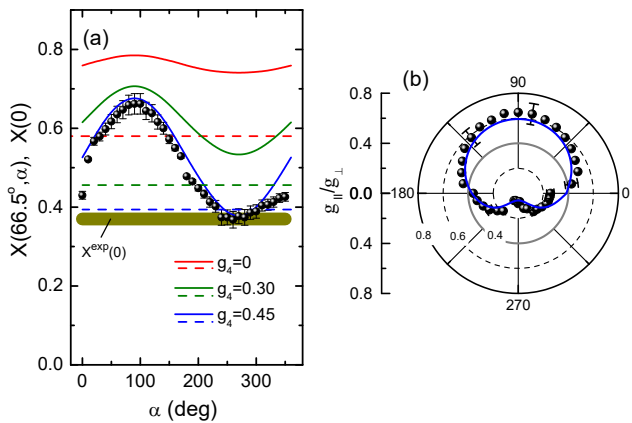}
\caption{(Color online) (a) -- The ratio of the spin-to-orbit splitting for $\theta=66.5^\circ$ plotted against the angle $\alpha$. Symbols are obtained from the $A(66.5^\circ, \alpha)/A(0)$  data shown in Fig.~\ref{F5}(b)  by using Eq.~(\ref{eq1}).  The solid curves and dashed lines are the  dependences $X(66.5^\circ,\alpha)$ and the $X(0)$ values, respectively, calculated for the different values of the parameter $g_4$. (b) -- The in-plane anisotropy of $\textsl{g}$-factor, $\textsl{g}_\parallel(\alpha)/\textsl{g}_\perp$. The circles are the data, the curve is the calculated dependence for $g_4=0.45$.
}
\label{F6}
\end{figure}

In  Fig.~\ref{F5}(a) we represent $d\rho_{xx}/dB_\perp$  as a function of $B_\perp$ for different $\alpha$ at $\theta=66.5^\circ$, where $\alpha$ is the angle between the  $B_\parallel$ direction and the axis [100] as shown in Fig.~\ref{F0}(b). As seen the positions of the oscillations are practically independent of $\alpha$, while  the oscillation amplitudes change dramatically, so that the oscillation phases jump on $\pi$ at certain angles: between $\alpha=0^\circ$ and $20^\circ$, and near $\alpha \approx 180^\circ$. The angle dependence $A(\theta=66.5^\circ, \alpha)/A(0)$ for $B_\perp=0.48$~T is plotted in Fig.~\ref{F5}(b). Using it  we find from Eq.~(\ref{eq1}) the ratio of the spin splitting to the orbital one $X(66.5^\circ,\alpha)$ which is plotted against $\alpha$ in Fig.~\ref{F6}(a).

The difference between  $X(66.5^\circ,\alpha)$ and X(0) can be interpreted qualitatively as the  contribution of in-plane magnetic field to the spin splitting. If we assume that the effective $\textsl{g}$-factor is still described by the expression Eq.~(\ref{eq2}) even for such a complex spectrum, we can obtain the $\textsl{g}$-factor anisotropy
$\textsl{g}_\parallel(\alpha)$/$\textsl{g}_\perp$ as follows
\begin{equation}
\frac{\textsl{g}_\parallel(\alpha)}{\textsl{g}_\perp}=
\frac{1}{\tan(\theta)}\sqrt{\left[\frac{X(\theta,\alpha)}{X(0)\cos(\theta)}\right]^2-1}.
\label{eq1001}
\end{equation}
The result of such a data treatment  is shown in Fig.~\ref{F6}(b). It is seen that the in-plane $\textsl{g}$-factor extremely anisotropic. So  the $\textsl{g}_\parallel/\textsl{g}_\perp$ value is close to $0.65$  for $\alpha\approx 90^\circ$, while at $\alpha\approx 270^\circ$ it is equal to zero within the experimental error.

Let us compare these data with the results of theoretical calculations. In Fig.~\ref{F6}(a), we present the results of calculations of $X(66.5^\circ,\alpha)$ and $X(0)$ carried out for this structure for the different $g_4$ values. It is evident that the calculations performed without taking into account the interface inversion asymmetry, i.e., with $g_4=0$, give  $X(66.5^\circ,\alpha)$ and $X(0)$ which significantly exceed the experimental data. Therewith the angle dependence $X(66.5^\circ,\alpha)$ is very weak, that means the weak anisotropy of the in-plane $\textsl{g}$-factor.

The possible reasons for the discrepancy between $X^\text{calc}(0)$ and $X^\text{exp}(0)$ were discussed  in Section~\ref{sec:th0}. The first reason is associated with a smaller value of $m^\text{exp}$ in comparison with  $m^\text{calc}$.  It was shown that taking  into account the electron mass renormalization can lead to a decrease in $X^\text{calc}(0)$ and thus to a good agreement with $X^\text{exp}(0)$. Our estimates show that such accounting for the renormalization  of $X^\text{calc}(\theta,\alpha)$  reduces  the value of $X^\text{calc}(66.5^\circ,\alpha)$ in $m^\text{calc}/m^\text{exp}$ times also, but it does not lead to an increase in anisotropy of in-plane $\textsl{g}$-factor.

The second reason is related to the interface inversion asymmetry. The calculation of $X(66.5^\circ,\alpha)$ with taking it into account for several  values of parameter $g_4$ are represented in Figs.~\ref{F6}(a) and \ref{F6}(b). As clearly seen, with an increase of the parameter $g_4$, the calculated $X^\text{calc}(0)$  and $X ^ \text{calc} (66.5^\circ,\alpha)$ values become closer to the experimental ones and  almost coincide with them when $g_4 = 0.45$~eV\,\AA.

The above comparison of theoretical calculations with experimental data shows that the taking into account interface inverse asymmetry  is necessary to obtain strong,  comparable with experiment, anisotropy of $X^\text{calc}(66.5^\circ,\alpha)$. This, however, does not mean that the mass renormalization does not play any role and therefore the value $g_4 =0.45$~eV\,\AA\, obtained when only IIA is taken into account should not be considered as determined  reliably.

The above results were obtained for the structure with $d=10\text{~nm}>d_c$ in which the conduction band is formed from  the heavy-hole states. The natural question arises: what role does the  inversion of the spectrum at $d>d_c$  play in the giant anisotropy of the in-plane $\textsl{g}$-factor? To elucidate this question we turn now to analysis of the data obtained for the structure 2 with normal band ordering.

\section{ In-plane $\textsl{g}$-factor anisotropy in the structure with	$d<d_c$
}
\label{secVI}

To find the in-plane $\textsl{g}$-factor anisotropy in the structure with $d<d_c$ all the measurements described  above were carried out for the structure 2 with $d=4.6$~nm (see Table~\ref{tab1}). Since the analysis of the results is analogous to that described above for the structure 1  we present the key data only.

\begin{figure}
\includegraphics[width=\linewidth,clip=true]{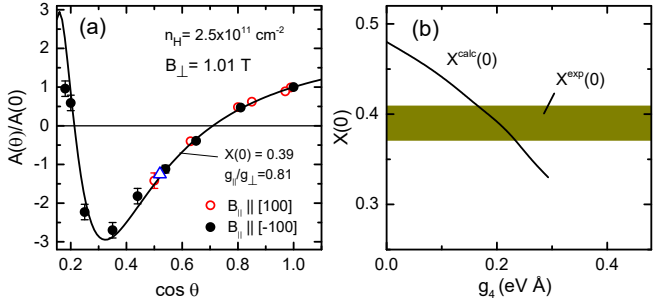}
\caption{(Color online) (a) -- The dependences of the normalized oscillation amplitude $A(\theta)/ A(0)$ at $B_\perp=1.01$~T. The symbols are the data, the curve is the dependences calculated from Eq.~(\ref{eq1}) for  $X(0)=0.39$ and $\textsl{g}_\parallel/\textsl{g}_\perp=0.81$. (b) -- The  ratio of the spin splitting to orbital one $X(0)$ plotted against the $g_4$ value.
}
\label{F7}
\end{figure}

The dependence of the oscillation amplitude on the angle $\theta$ obtained in the first configuration (the rotation axis lies in the 2D plane and is perpendicular to the magnetic field) is presented in Fig.~\ref{F7}(a). The data shown by the open and solid circles are obtained when the in-plane $B$ component appears in the direction $[100]$ and $[\bar{1}00]$, respectively. One can see that the angle dependences of relative amplitude   $A(\theta)/A(0)$ for both directions coincide with each other within the experimental error. The formula Eq.~(\ref{eq1}) describes well both sets of the data with the parameters $X(0)=0.39$ and $\textsl{g}_\parallel/\textsl{g}_\perp=0.81$. The error is estimated as $\pm 0.02$ for both parameters.

The theoretical  value of $X(0)$ calculated with neglecting  the interface inversion asymmetry is equal to $X^\text{calc}(0)=0.46$ that is greater than $0.39$ found experimentally. As discussed in Section~\ref{sec:th0} the two factors, renormalization of the effective mass and IIA, can be the reasons for this discrepancy. The second factor can be excluded because, contrary to the structure 1 with $d=10$~nm, the effective  mass measured for structure 2 at $n=2.3\times10^{11}$~cm$^{-2}$ practically coincides with the calculated one:  $m^\text{exp}=(0.0225\pm 0.002)\,m_0$, $m^\text{calc}=0.0222\,m_0$. As for the role of the first factor,  $X^\text{calc}(0)$ coincides with $X^\text{exp}(0)$ when $g_4\simeq 0.2$~eV\,\AA\, [see Fig.~\ref{F7}(b)].  It should be noted that this  value is seemingly less than that for structure 1 with $d=10$~nm and strongly less than $g_4$ obtained in Ref.~{\cite{Minkov17}} where the valence band spectrum is investigated.

\begin{figure}
\includegraphics[width=\linewidth,clip=true]{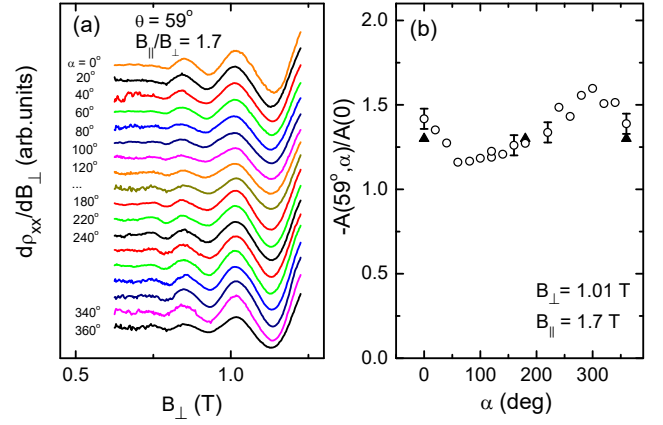}
\caption{(Color online) (a) -- The $B_\perp$ dependences of  $d\rho_{xx}/d B_\perp$ for  the different   angles $\alpha$  measured for a fixed angle $\theta=59^\circ$. (b) -- The normalized oscillation amplitudes $A(59^\circ, \alpha)/A(0)$ plotted against the  angle $\alpha$ for $B_\perp=1.01$~T. The triangle shows  $A(59^\circ, 0)/A(0)$  obtained in the  first configuration [see Fig.~\ref{F7}(a)]. Structure 2.
}
\label{F8}
\end{figure}

\begin{figure}
\includegraphics[width=\linewidth,clip=true]{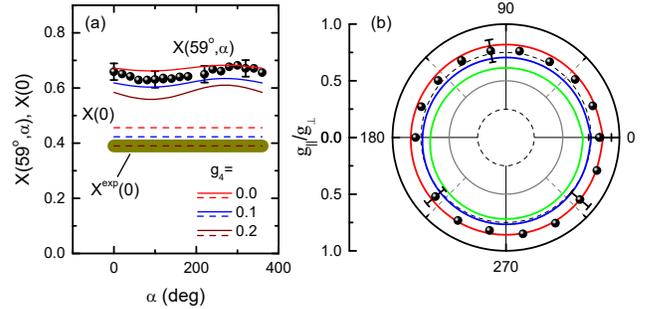}
\caption{(Color online) (a) -- The ratio of the spin-to-orbit splitting for $\theta=59^\circ$ plotted against the angle $\alpha$. Symbols are obtained from the $A(59^\circ, \alpha)/A(0)$ data shown in Fig.~\ref{F8}(b)  by using Eq.~(\ref{eq1}).   The solid curves and dashed lines are the  dependences $X(59^\circ,\alpha)$ and $X(0)$, respectively, calculated for the different $g_4$ values.   (b) -- The   anisotropy of in-plane  $\textsl{g}$-factor. The circles are the data, the curves are the theoretical dependences calculated  with the same $g_4$ values as in the panel (a). }
\label{F9}

\end{figure}

Let us inspect the data measured in the second configuration. The oscillation curves measured at $\theta=59^\circ$ for different angles $\alpha$ and the $\alpha$ dependence of the normalized oscillation amplitude at $B_\perp=1.01$~T  are shown in Figs.~\ref{F8}(a) and \ref{F8}(b), respectively. As seen the $\alpha$ dependence of the oscillation amplitude for this structure is much weaker than that for the structure 1 (see Fig.~\ref{F5}) that points to the relatively weak anisotropy of the in-plane $\textsl{g}$-factor directly. The last is illustrated by Fig.~\ref{F9} in which  the dependences $X(59^\circ,\alpha)$ and $\textsl{g}_\parallel(\alpha)/\textsl{g}_\perp$ obtained from the data presented in Fig.~\ref{F8} are shown. It is  seen that  the in-plane $\textsl{g}$-factor anisotropy is really  weak  in the structure 2 with $d<d_c$. In the same figure, the results of theoretical calculation are shown. One can see that the satisfactory agreement between theory and experiment for the angle dependences $X(59^\circ, \alpha)$  is achieved with $g_4=(0.1\pm 0.05)$~eV\,\AA.

\section{ Conclusion}
\label{sec:conc}

The Shubnikov-de Haas effect in the conduction band was investigated in the (013)-Hg$_{1-x}$Cd$_x$Te/HgTe quantum wells  both with normal and ``inverted'' energy spectrum. Analyzing the oscillations measured for the different orientations of magnetic field relative to the QW plane and crystallographic directions we have obtained  the anisotropy of the ratio of the spin splitting to the orbital one. The data relevant to the QWs with normal and ``inverted'' energy spectra differ significantly.

For the QWs with ``inverted'' spectrum, it has been shown that this ratio is strongly dependent both on the orientation of the magnetic field relative to the QW plane and on the orientation of the in-plane component of magnetic field relative to crystallographic axes laying in the QW plane. As for the QW with normal spectrum,  this ratio being essentially dependent upon the angle between the magnetic field and normal to the QW-plane reveals only weak anisotropy in the QW plane.

To interpret the data obtained, the Landau levels in the tilted magnetic field have been calculated within the framework of the four band \emph{kP}-model. It has been shown that the experiment results can be quantitatively described only with taking into account interface inversion asymmetry. This allows us  to estimate the value of the parameter $g_4$ responsible for the IIA  contribution. It is several times smaller  than that obtained for the valence band in Ref.~\cite{Minkov17}. In our opinion, this could indicate that the approximation in which the sole parameter $g_4$ is responsible for the IIA  contribution to the spectra both of conduction and valence bands is not good enough and the more accurate approach is needed.

\begin{acknowledgments}
We are grateful to E.~L.~Ivchneko  for useful discussions.
The work has been supported in part by the Russian Foundation for Basic
Research (Grant No. 18-02-00050), by  Act 211 Government of the Russian Federation, agreement No.~02.A03.21.0006,  by  the Ministry of Education and Science of the Russian Federation under Project No. 3.9534.2017/8.9, and by the FASO of Russia (theme ``Electron'' No. 01201463326).
\end{acknowledgments}


%

\end{document}